# NEW CALIBRATIONS OF PULSATIONAL ABSOLUTE MAGNITUDES OF FIELD RR LYRAE STARS USING REVISED DEPENDENCIES OF TEMPERATURES, MASSES, AND PERIODS ON METALLICITY

Allan Sandage

The Observatories of the Carnegie Institution of Washington, 813 Santa Barbara Street,

Pasadena, CA 91101, USA

## ABSTRACT

The pulsation method to estimate the absolute magnitudes of RR Lyrae stars is updated with new data on field-star metallicities by Layden, a new calibration of the $(B-V)_o$-temperature correlation from recent atmospheric models by Bell and Tripicco, and new mass estimates by Bono et al. New linear and non-linear calibrations of $M_V(RR) = f([Fe/H])$ are derived that depend on the nature of the adopted envelope in a diagram of log period versus metallicity fitted to the shortest period field variables in each metallicity range, together with the stated assumptions on the colour of the stars at that envelope. These new calibrations are compared with a recent non-linear calibration by Caputo et al.

The theoretical luminosity zero points for each of the three new calibrations derived here agree with each other to within 0.1 mag over the metallicity range of $-1.0 > [Fe/H] > -2.0$. Comparison with the empirical absolute magnitude calibration of $M_V(RR) = +0.52$ at $[Fe/H] = -1.5$ by Clementini et al. from RR Lyraes in LMC also shows agreement at the 0.05 mag level at the stated metallicity. Our final compromise pulsational calibration is,
$$M_V(RR) = 0.82 + 0.20 \, ([Fe/H])$$
with the caveat that the true relation is likely to be non-linear at the 0.1 mag level for $[Fe/H]$ values between -0.5 and -2.0, and hence with variable $dM_V/d([Fe/H])$ slopes that cover the range from 0 to 0.6 for different metallicities. From the manner of the setting of the zero point of this equation using the Clementini et al. LMC data, this calibration refers to the average level of the horizontal branch as it is affected by luminosity evolution away from the age zero horizontal branch.

1. INTRODUCTION

There are six methods currently in use for calibrating the absolute magnitudes of RR lyrae stars, each with their merits and each with various difficulties and systematic uncertainties. They are (1) main sequence fittings of colour-magnitude diagrams of globular clusters to obtain distances of clusters that contain RR Lyrae variables, (2) statistical parallaxes of field variables determined by equating radial velocities with the tau components of the proper motions for field RR Lyrae variables, (3) the Baade-Becker-Wesslink (BBW) method of combining pulsational velocities with measured surface brightnesses over the cycle, (4) use of a theoretical pulsation equation that relates luminosity to observed values of period, mass, and temperature at various places in the instability strip, (5) comparison of the observed apparent magnitude of the tip of the red giant branch with theoretical models that give the absolute magnitude for the helium flash that terminates the branch, and (6) comparison of RR Lyrae apparent magnitudes with classical Cepheids in external galaxies where both types of variables are seen together. The expanding literature on each of these methods, complete to the mid 1990s, has been reviewed by Smith (1995).

The purpose of this paper is to update the pulsation method that was used with the cluster and field star data a decade ago (Sandage 1993a,b, hereafter S93a,b). Five developments since the 1993 analysis show the need to rediscuss the derived luminositites using the method.

(1) A new determination of the metallicities of a large sample of field RR Lyraes has been made by Layden (1994) on the metallicity scale of Zinn and West (1984). This homogenizes the data used in S93a,b where the metallicity scale of Zinn and West was used for the globular clusters, but where the field star data were analyzed on the metallicity scale of Blanco (1992) that itself was based on the scale of Butler (1975). (2) A new determination of the envelope at the short period edge of the diagram by Preston (1959) that relates log period and metallicity of field RR Lyraes suggests a non-linear period-metallicity envelope that replaces the linear approximation in S93a. Suggestions can also be advanced why this envelope may not define the blue edge of the instability strip as was assumed in S93a, but may be distorted by the Oosterhoff (1939, 1944) period dichotomy as it is related to metallicity (Arp 1955, Preston 1959). (3) New calibrations of the temperature scale as functions of $(B-V)_o$, metallicity, and surface gravity have been made by Kovacs & Walker (1999), Sandage, Bell & Trippico (1999, hereafter SBT), and Sekiguchi & Fukugita (2000), among others, permitting tests of the sensitivity of the predicted pulsation absolute magnitudes for different temperature-colour scales. (4) Bono et al. (1997) have computed new age zero horizontal branch (AZHB) models, permitting comparison with the mass-metallicity relation of Dorman (1992) used in S93a. (5) An analysis by Caputo et al. (2000) that gives a non-linear luminosity-metallicity relation, requires discussion.

We address each of these five points in this paper. The organization is as follows. A new non-linear (parabolic) envelope to the log P-metallicity correlation is derived in section 2, replacing the linear envelope used in S93a,b. Here we use Layden's (1994) metallicities for field RR Lyraes. This new envelope may or may not define the period at the blue edge of the instability strip for fundamental-mode variables more metal poor than -1.8. The distribution of stars along the AZHB (i.e. its morphology), plus luminosity evolution away from it, may be the cause of the indecisiveness. A second envelope to the log P-metallicity correlation is derived, taking this possibility into account. A third period-metallicity correlation is derived from the mid point ridge-line put through the Layden period-metallicity field star data.

Section 3 summarizes the new calibrations of the temperature and mass of RR Lyraes as functions of metallicity.

In section 4 we combine the results of sections 2 and 3 with the pulsation equation of van Albada & Baker (1973) to derive three new calibrations of $M_V(RR,[Fe/H])$, [eqs. 12, 13, 14], depending on which log P, [Fe/H] relation is used from section 2. These calibrations replace the linear calibration of

$$M_V(RR) = 0.94 + 0.30 \, ([Fe/H]), \qquad (1)$$

derived in S93b.

In section 5 we compare the predictions from section 4 with the non-linear $M_V([Fe/H])$ calibration of Caputo et al. (2000) that is based on their calculations of the position of the edges of the instability strip in the HR diagram. Section 6 replaces our theoretical absolute magnitude zero points in section 4 with an empirical calibration derived from the observed apparent magnitudes of RR Lyraes in the LMC at its independently known distance. Section 7 summarizes some of the literature values of the $dM_V/d([Fe/H])$ slope compared with the values we derive here. Section 8 is a summary.

## 2. THREE POSSIBLE CORRELATIONS OF PERIOD AND METALLICITY DERIVED FROM LAYDEN'S FIELD STAR DATA

*2.1. The Oosterhoff period dichotomy in globular cluster RR Lyrae variables and the period-metallicity continuum for the field variables*

A correlation between Oosterhoff's mean periods and metallicity for RR Lyraes in clusters was discovered by Arp (1955) in his demonstration that Oosterhoff's (1939, 1944) period separation into two discrete groups is also a separation by metal abundance. Confirmation was made in an important advance by Kinman (1959).

A model was proposed (Sandage 1958; Sandage, Katem & Sandage 1981, hereafter SKS) that provides an explanation of the period ratio dependence on metallicity for the two discrete period groups. It posited that the horizontal branch luminosity of metal poor clusters is brighter than that of more metal rich clusters. If so, the radii of the stars *at a given position* in the strip (i.e. at a given temperature) for the metal poor stars will be larger than at the same strip position in the metal rich clusters (at fixed mass) because of the luminosity difference. Providing that the mass of RR Lyraes is not a strong function of luminosity, the pulsation period of stars at the brighter luminosity but at the same temperature, will, necessarily, be longer than the period of those at the fainter level, based on the Ritter pulsation relation between period and the inverse square root of the density.

This model predicts that log period *shifts* (differences in log P, and ratios in straight periods) that depend on the metallicity will exist in the period-colour, period-amplitude, period-rise times, and period-$\phi_{31}$ (Fourier) relations for the variables. These relations soon became well observed from cluster to cluster where there are metallicity differences (SKS 1981; Sandage 1981a,b; 1990a,b; 2004; hereafter S81a etc.).

However, Oosterhoff's division into two *discrete* period groups must have a more complicated explanation than is given by this first-order model. The period *dichotomy* has variously been attributed to the way the horizontal branch morphology changes with metallicity as the branch threads in and out of the instability strip, changing directions of the evolutionary tracks in the HR diagram away from the age zero horizontal branch as a function of mass. This explanation has been given in various forms and in various detail by Renzini (1983), Castellani (1983), Gratton, Tornambe & Ortolani (1986), Lee (1989, 1990, 1991 1992), Lee, Demarque & Zinn (1990), Sandage (1990b), Bencivenni et al. (1991), Caputo & De Santis (1992), Catelan (1992), Caputo et al. (1993), Caputo (1997), and we suspect others. We return to this point below.

Preston (1959) later showed that the RR Lyrae variables in the general field form a *continuum* in a log P-metallicity diagram, rather than a dichotomy. The same continuum was later shown in the globular clusters when the cluster sample was increased (S82a; S93a,b), (cf. also various period-metallicity diagrams for dwarf elliptical companions of the Galaxy such as by Siegel & Majewski 2000, their Fig. 6).

The continuum relation in the Galactic globular clusters emerged when the periods of the cluster RR Lyrae variables were plotted on horizontal lines in a period, metallicity diagram, showing the period distribution for a given cluster. When these lines are stacked vertically by metallicity, a continuous cluster period-metallicity diagram is produced (S93a, Figs. 1-11). However, a discontinuity appeared in the distribution beginning near [Fe/H] = -1.9. The shortest period variables in any cluster are displaced toward longer periods than is predicted by a linear envelope fitted at higher metallicitues to the short period edge of the distribution. From this, it was clear that the Oosterhoff dichotomy into two period groups could be inferred from this discontinuity (S93a, Fig. 1).

Nevertheless, the discontinuity was neglected in S93a, and a single "best fitting" linear envelope, mentioned earlier, was fitted to the data (Fig. 10 of S93a) with the equation of,

$$\log P = -0.117 [Fe/H] - 0.526. \qquad (2)$$

*2.2. A parabolic envelope fit to the period-metallicity field star data*

A better fit to the envelope data in S93a would have been two straight lines with different slopes, joined at [Fe/H] = -1.8, with the equations $\log P = -0.084 ([Fe/H]) - 0.476$ for metallicites more metal rich than [Fe/H] = -1.8 and $\log P = -0.200 ([Fe/H]) - 0.685$ for [Fe/H] < -1.8. Had this linear double-envelope representation been used in the pulsation equation (S93b, Eq. 3'), together with the temperature-metallicity and mass-metallicity relations also used in S93b (eqs. 5 and 7 there), a *two-segment* absolute magnitude-metallicity relation would have been predicted. The equations would have been

$$M_V = 0.87 + 0.19 [Fe/H],$$

for [Fe/H] > -1.8 and,

$$M_V = 1.50 + 0.53 \, [Fe/H],$$

for [Fe/H] < -1.8, using $M_{bol}(sun) = 4.75$ and the bolometric correction of $0.06 + 0.06 \, [Fe/H]$ from equation (9) in S93b. If so, the absolute luminosities would be brighter for more metal poor variables than the linear $M_V = f([Fe/H])$ relation of equation (1). The slope, $dM_V/d[Fe/H]$, would vary with metallicity between 0.19 and 0.53. Caputo et al. (2000) emphasized this possibility of a variable slope with metallicity (but with the opposite sense), both empirically and by citing various theoretical predictions in a number of current AZHB models (see below).

The correlation in Fig. 1 of log period and metallicity for *field* RR Lyraes shows the same non-linear trend as the two-line fit of the envelope just mentioned. Figure 1 here is similar to Fig. 1 of Layden (1995) but we use log period rather than period and have plotted [Fe/H] as ordinate rather then abscissa.

The parabolic envelope in Fig. 1 has the equation,

$$\log P = -0.452 + 0.033 \, ([Fe/H])^2 \tag{3}$$

We have dropped TV Lib (log P = -0.569, [Fe/H] = -0.27), V784 Oph (-0.425, -1.80), and SV Vol (-0.422, -2.18) from the diagram. Four others, AA Aql (-0.215, -0.58), SS Gru, (-0.310, -2.35), V445 Oph (-0.410, -0.23), and WY Vir (-2.15, -2.13) are slightly outside the line.

*2.3. A linear fit that includes the four outriders*

If we include the four aforementioned stars AA Aql, SS Gru, V455 Oph, and WY Vir, forcing a straight line through the first two, the linear envelope line shown in Fig. 2 obtains. It neglects the hole in the distribution for [Fe/H] < -2.1 and log P between -0.3 and -0.25. This is the feature that produces the Oosterhoff period *dichotomy* in clusters. The line in Fig. 2 has the equation,

$$\log P = -0.484 - 0.074 \, [Fe/H], \tag{4}$$

to be used in section 4 as an alternative to equation (3).

*2.4. A ridge line through the mean points of the period-metallicity correlation*

Because neither equation (3) or (4) can be decisively proved to describe the period-metallicity distribution at the fundamental blue edge (because of the discontinuity) we posit that it may be more reasonable to work with the "ridge line" of the Fig. 1 correlation. Points on this line, with the average log P values calculated in narrow bins of [Fe/H], are given in Table 1. The least squares equation of the line is,

$$\log P = -0.416 - 0.098 \, ([Fe/H]), \tag{5}$$

to be used as an alternative to both equations (3) and (4) in section 4.

Also listed in Table 1 are the mean $(B-V)_o$ colours within each bin of metallicity, calculated from the photoelectric colours listed in column 5 of Nikolov, Buchantsova, and Frolov (1984, hereafter the Sophia Catalog) with a reddening correction based on the absorptions (divided by 3) measured by Layden (1994). These colour data are needed in section 3.3.

## 3. NEW TEMPERATURE AND MASS RELATIONS FOR RR LYRAES AT VARIOUS PLACES IN THE INSTABILITY STRIP AS FUNCTIONS OF METALLICITY

New calibrations of masses, effective temperatures at various $(B-V)_o$ colours, and metallicities have appeared since S93b. We compare them in this section.

*3.1 Four independent calibrations of effective temperatures for stars near the instability strip*

A summary of various colour-temperature calibrations has been made by Sekiguchi and Fukugita (2000). They cite a number of independent calibrations in the post 1980 literature; papers prior to 1979 are subsumed in the earlier review by Bohm-Vitense (1980). In deriving their own colour-temperature calibration, they rely on the advance made by Di Benedetto (1998) who used the measured angular diameters of giants and dwarfs and correlated the derived temperatures, therefrom, with known $(V-R)_o$ colours on the Cape Cousins photometric system. Sekiguchi and Fukugita then transferred Di Benedetto's $(V-R)_o$-temperature relation to a larger sample of stars with known $(B-V)_o$ colours, deriving an equation relating $(B-V)_o$, [Fe/H], and log g over the spectral range from F to K stars, metallicities between [Fe/H] of 0.0 and -2.2, and log g between 4.5 and 3.0. The virtue of their procedure is its minimum dependence on stellar atmospheric models, but rather the use of observed angular diameters put directly in the defining equation for effective temperature.

Dorman (1992) tabulates $(B-V)_o$ colours for a range of effective temperatures, metallicites, and surface gravities for his oxygen enhanced horizontal branch models. His $(B-V)_o$ colour scale for metallicities between -0.47 > [Fe/H] > -2.26 is that of VandenBerg (1992), and Bergbusch and VandenBerg (1992), which is based on the calibrations of Bell and Gustafsson (1978), Kurucz (1979), and VandenBerg and Bell (1985).

Kovacs and Walker (1999) give an equation that relates $T_e$, $(B-V)_o$, [Fe/H], and surface gravity, derived from theoretical static atmospheric models by Castellani, Gratton, and Kurucz (1997), where the zero points of the calibration are based on the ATLAS code of Kurucz (1993).

New atmospheric models by Bell and Trippico (SBT 1999) produces a grid of colour transformations for different metallicites ranging from [Fe/H] of +0.3 to -1.66, surface gravities that embrace dwarfs, giants, and supergiants, and two values of the mean atmospheric turbulent velocity of 1.7 km s$^{-1}$ and 5.0 km s$^{-1}$. Details of the input provenances for the Bell/Tripicco models are in SBT.

*3.2 Comparisons of the four colour-temperature calibrations*

Comparisons of these four independent colour-temperature calibrations show remarkable agreement to generally within $\Delta \log T_e = 0.01$ over the colour range of $0.2 < (B-V)_o < 0.7$ mag.

Because the temperature enters the pulsation equation (11) with such a high power, the comparison is important in the assessment in section 4 of the reliability of both the zero point of the predicted RR Lyrae luminosity and its dependence on metallicity.

Table 2 sets out the details of the comparison of the four calibrations as functions of $(B-V)_o$ and metallicity. Log g was assumed to be 3.0 throughout the comparisons. The dependence of $T_e$ on $(B-V)_o$ has only a slight surface gravity dependence (SBT, Table 6), so exact values of log g are not important.

The sense of the residuals in Table 2 is SBT minus others (using 1.7 kms$^{-1}$ for the atmospheric turbulence in SBT). SF refers to the calibrations of Sekiguchi and Fukugita (2000). KW refers to the calibration of Kovacs and Walker (1999). Positive residuals mean that the SBT temperatures are hotter than the comparison values.

The following differences are seen from Table 2. The SF temperatures are cooler than those of SBT for $(B-V)_o$ bluer than 0.3, [Fe/H] < -0.5, by residuals than reach $\Delta \log T_e = 0.010$, or temperature differences of 150$^o$ K at $T_e = 6500$ K$^o$. However, at the crucial colour of $(B-V)_o = 0.24$ (see the next subsection), the SBT and SF temperatures are nearly identical for [Fe/H] = -0.5. Thereafter, for redder $(B-V)_o$ colours, the SF temperatures become hotter than SBT for lower metallicities. Parallel statements can be made (but mind the sense) for the KW and Dorman scales by again reading Table 2.

Table 3 shows that the KW scale has the same slope of $d\log T_e/d[Fe/H] = 0.07$ for the colour-temperature relation for all [Fe/H], whereas the other scales show the variable slopes with varying [Fe/H] that is expected from the increased blanketing and backwarming with increasing colour (eg. Sandage & Eggen 1959), i.e. the slopes of the blanketing lines in the two-colour diagram depend on colour, as in Wildey et al. (1962).

*3.4 Temperatures at the blue fundamental edge of the instability strip and at the midpoint of the period-metallicity relation*

In the pulsation equation (11), set out below, we need the temperature, either at the fundamental blue edge using equations (3) and (4), or at the midpoint of the type ab instability strip using equation (5) and the colours in Table 1.

The reddening-corrected colour of the blue fundamental edge is taken to be $(B-V)_o = 0.24$ for all [Fe/H] values. This colour was also adopted in the S93b analysis, justified in Fig. 1 there. It is based on the unreddened colour of field RR Lyraes. In S93b we used Blanco's reddening values applied to the observed photoelectric values in the Sophia Cataloge. In the present paper we use Layden's (1994) absorption values, divided by 3, applied to the photoelectric colours in the Sophia catalog.

This adopted blue-edge colour is also the colour of the bluest type ab variables in M3, M15 and in six other clusters for which we earlier had photoelectrically measured photometry (Sandage 1970, S90a). The more recent precision CCD photometry of a number of clusters have confirmed this value to within 0.02 mag, (eg. Reid 1996, and Caputo et al. 1999 for M5; Corwin and Carney 2001 for M3; Walker and Nemec 1996 for IC 4499; Walker 1992, and recently many others in an exploding literature), but to be sure, depending on the E(B-V) adopted for the individual clusters.

However, it must be emphasized that even though the colour at the blue edge is adopted to be constant for variable metallicity, nevertheless the *temperature* of the blue edge *does* depend on [Fe/H] because the colour-temperature relation is metallicicty dependent, shown in Table 3. The sense is that the blue-edge temperature becomes cooler as the metallicity decreases, at constant colour, at the rate given in the Table, entirely consistent with calculations of the differences in Fraunhofer blanketing as a function of metallicity.

It is, of course, well established that the edges of the instability strip in the HR diagram indeed do slope toward cooler temperatures for higher luminosities at the rate of dlog T/d$M_V$ = 0.030 (eg. many references in the standard literature starting with Iben and Tuggle, 1972), (cf. also Figs. 1-4 in Sandage, Dietheim, & Tammann, 1994).

Hence, if the AZHBs of metal poor clusters is brighter than that for metal rich clusters, their blue-edge strip temperatures should be cooler than those for the metal rich edges even through their blue-edge colour is taken to be the same. For example, if the luminosity difference for the Oosterhoff effect is 0.25 mag for a metallicity difference of $\Delta$[Fe/H] = 0.7 (the difference between M3 and M15), the temperature difference of the blue edge (at constant colour) is expected to be $\Delta \log T_e$ = (0.030)(0.25) = 0.008, the metal poor blue edge being cooler. Therefore, the slope of the temperature-metallicity relation (again at constant colour) due to the sloping edges of the instability strip should be dlog T/d[Fe/H] = 0.008/0.7 = 0.011, for this metallicity difference. *This is nearly identical to the entry in Table 3 for the SBT temperature scale at constant $(B-V)_o$ = 0.24.*

This near perfect agreement shows the coherence of a model where (1) the colour at the blue edge is the same for all metallicities, (2) an increased luminosity of metal poor stars exists as required in the explanation used here for the Oosterhoff period ratio-metallicity effect at the rate of ~0.25 mag per 0.7 differences in [Fe/H], and (3) the fact that the edge of the instability strip slopes toward cooler temperatures at higher luminosities.

Consider, then, the temperature at a colour of $(B-V)_o$ = 0.24 for different metallicites, based on the temperature scales cited in this section. Comparing three of the four scales discussed above (the constant slope to the temperature-colour scale for different metallicities in KW is suspect because it should vary due to a blanketing slope dependence on [Fe/H]), we derive linear approximations to the temperature at the blue edge fixed at $(B-V)_o$ = 0.24 for all [Fe/H] of

$$\log T_e = 3.872 + 0.010 \text{ [Fe/H]}, \tag{6}$$

for the SBT scale,

$$\log T_e = 3.879 + 0.018 \text{ [Fe/H]}, \tag{7}$$

for the SF scale, and

$$\log T_e = 3.857 + 0.007 \text{ [Fe/H]}, \tag{8}$$

for Dorman's scale. More accurate non-linear log $T_e$([Fe/H]) relations can be derived from the details published for these various scales over the metallicity range from 0.0 to -2.0, but the linear versions in equations (6) to (8) are sufficiently accurate to illustrate the problems we address here.

Recall that we used log $T_e$ = 3.872 + 0.012 [Fe/H] in S93b at the fundamental blue edge, similar to equation (6) here.

At $(B-V)_o = 0.33$ for the colour of the mid-point ridge line of the log period-[Fe/H] relation of Fig. 2, the SBT equation for the temperature-[Fe/H] relation is

$$\log T_e = 3.844 + 0.016 \text{ [Fe/H]}, \quad (9)$$

to be used in Sec 4.

The four calibrations we are discussing have a colour-temperature slope of $d\log T_e/d(B-V) = 0.4$ at $(B-V)_o = 0.30$ at [Fe/H] = -1.0. Hence, an error of 0.02 mag in the colour of the blue edge translates to an error of 0.008 in $\log T_e$, to be used later (Sec. 4.2).

*3.4 The adopted mass-metallicity relation*

Dorman's 1992 AZHB models were computed using enhanced oxygen abundances and modern opacity tables. Dorman predicted a mass variation with metallicity of $\log M = -0.059 \text{ ([Fe/H])} - 0.288$, which we adopted in S93b. More recently, Bono et al. (1997) have calculated new AZHB models also with updated opacities and, as in Dorman, with oxygen enhancements. The result is,

$$\log \text{Mass (solar units)} = -0.066 \text{ ([Fe/H])} - 0.283, \quad (10)$$

which we adopt for the calculations in section 4. The consequences of using this equation rather than Dorman's for the mass is only 0.01 mag in the calculated absolute magnitude zero point, and 0.014 mag in the sensitivity of the absolute magnitudes to [Fe/H] (section 4) over the whole of the relevant metallicity range.

Although equation (10) from Bono et al., as with Dorman's earlier equation, is from *theoretical* AZHB models, the two equations agree well with the empirically determined mass for various metallicities derived from data on double mode pulsators based on the observed period ratios, $P_o/P_1$, of the fundamental to the first harmonic via the Petersen (1973) diagram. Hence, the masses here can be said to be determined "observationly".

## 4. PREDICTED LUMINOSITY-METALLICITY CALIBRATIONS USING THE ENVELOPE LINES IN SECTION 2 AND THE TEMPERATURE AND MASS RELATIONS IN SECTION 3

The pulsation calculations of van Albada and Baker (1973) have been repeated by many authors over the past 25 years using a variety of metal and helium abundances and with various opacity values. The result is that the vA/B equation remains a surprisingly excellent description using the assumed variation of He abundance with [Fe/H], i.e. there is no appreciable metallicity dependence of the coefficients of its various terms on chemical composition for reasons argued in an extended discussion by SBT (1999). For consistency with S93b, we continue to use this equation, which is,

$$\log L = 1.19 \log P + 0.81 \log \text{Mass} + 4.143 \log T_e - 13.687, \quad (11)$$

independent of metal and helium abundance. Of course, the *positions* of the edges of the instability strip itself in the HR diagram *are* dependent on chemical composition, but we do not need these theoretical $M_{bol}$, log P strip positions to predict the pulsational luminosity from equation (11) because we use the *observed* period/[Fe/H], mass/[Fe/H], and temperature/[Fe/H] relations to eliminate the period, mass, and temperature in equation (11), leaving only the dependence of $M_{bol}$ on [Fe/H].

## 4.1. $M_V$(RR) using the boundary line in Fig. 1

Substituting the period/metallicity envelope from equation (3) into the pulsation equation (11) together with equation (6) for the SBT temperature/metallicity relation at $(B-V)_o = 0.24$, equation (10) for the masses, and using a new bolometric correction derived from Table 6 of SBT as $BC = 0.060 + 0.069$ [Fe/H] at $(B-V)_o = 0.24$ (with $M_V$ defined as $M_{bol}$ minus BC), and adopting the bolometric absolute magnitude of the sun to be +4.75, gives the absolute magnitude as,

$$M_V(RR) = 0.718 - 0.039 ([Fe/H]) - 0.098 ([Fe/H])^2, \qquad (12)$$

for stars on the Fig. 1 envelope. The zero point is solely theoretical, depending on the constant in equation (11) and the adopted zero point constants in equations (3), (6), and (10). Equation (12) is shown as the line marked number 1 in Fig. 3. The other lines are discussed below.

Equation (12) is non-linear. It predicts a brightening toward lower metallicities compared with a linear relation. Moreover, the sense of the curvature is opposite that of Caputo et al. (2000), discussed in the next section. Clearly, the slope, $dM_V/d([Fe/H])$, so sought after in the literature, varies with [Fe/H]. Equation (11) accommodates slopes between 0.0 and 0.45 as [Fe/H] varies between 0.0 and -2.5.

## 4.2. Pulsation $M_V$(RR) using the alternative linear blue edge period-metallicity relation of equation (4)

The linear envelope given by equation (4) which accommodates AA Aql and SS Gru in Fig. 2, is as extreme a fit to the log period-metallicity correlation as can be made without violating the data. In particular, no curvature toward the left in the log P-metallicity plane can be accommodated, which we show in section 6 and Fig. 3 *would* be required to account for the Caputo et al. (2000) calibration (section 5) if the temperature of the blue edge is that given by equation (6).

With the linear envelope of equation (4) whose slope is $dlogP/d[Fe/H] = -0.074$, put in the pulsation equation (11), and again using equation (6) for the temperature, and equation (10) for the mass at the blue fundamental edge, gives the absolute magnitude calibration as

$$M_V(RR) = 0.82 + 0.18 ([Fe/H]), \qquad (13).$$

shown as the line marked number 2 in Figure 3.

Remarkably, for all [Fe/H between 0 and -2.5, equation (13) is everywhere within 0.1 mag of the *empirical* calibration of Carretta et al. (2000) whose adopted relation is $M_V$(RR) =

0.89 + 0.18 ([Fe/H]). Equation (13) is also everywhere within 0.02 mag of Fernley's (1993) empirical calibration of 0.84 + 0.19 ([Fe/H]) for all [Fe/H] > -2.5.

*4.3. Pulsation $M_V(RR)$ from data at the midpoint of the type ab instability strip*

On the supposition that the midpoint period-metallicity relation, shown as the dashed line in Fig. 2, may avoid here a putative Oosterhoff dichotomy effect, we calculate $M_V(RR)$ at the midpoint using equation (5) for the period-metallicity relation, equation (10) again for the masses, but now equation (9) for the temperature at an assumed midpoint colour of $(B-V)_o = 0.33$, based on Table 1. These, put into the pulsation equation (11) using the bolometric correction by interpolation in Table 6 of SBT as BC = 0.050 +0.092 ([Fe/H]) at $(B-V)_o = 0.33$, give,

$$M_V(RR, \text{midpoint}) = 0.91 + 0.17 ([Fe/H]). \quad (14)$$

This is 0.1 mag fainter than equation (13) at the blue edge. However, all the zero points in equations (12), (13), and (14) are sensitive to the zero points of the temperature scales due to the large coefficient to the temperature term in the pulsation equation. From the colour-temperature gradient of $d\log T_e/d(B-V)_o = 0.4$ (Sec 3.3), a change of the mid-point colour from 0.33 to 0.31 makes the temperature hotter by $\Delta \log T_e = 0.008$. Then, from equation (11), the predicted absolute magnitude is brighter by (4.143)(0.008)(2.5) = 0.08 mag, which would nearly reconcile the 0.1 mag difference between equations (13) and (14).

5. COMPARISON WITH THE NON-LINEAR CALIBRATION OF CAPUTO ET AL. (2000)

Caputo et al. (2000), updating an earlier application (Caputo 1997) of the pulsation method, have used the pulsation equation in a different way then we have done here and in S93a,b. They combine pulsation calculations for the periods with new theoretical models of the period-luminosity relations at the blue edge of the first harmonic instability strip. The new models, set out in a series of papers cited by CCMR, include those of Bono and Stellingwerf (1994), Bono, Caputo, and Marconi (1995), and Bono et al. (1997) using the new opacities of Iglesias and Rogers (1996).

CCMR then construct a series of loci from their Table 1 to make a diagram of $M_V$ vs. log P, binned by metallicity, for the blue edge of the first harmonic and the red edge of fundamental mode pulsators. These loci are straight lines in a $M_V$-log P diagram (their Fig. 1).

They then fit the observational data on *apparent* magnitudes and log P for first harmonic (Bailey type c) RR Lyraes in a number of clusters, themselves binned by metallicity, adjusting the cluster distances by vertical displacements to the first harmonic blue edge in the diagrams until the observed shortest period variables in a given cluster satisfy the position of the theoretical $M_V$, log P loci for the appropriate metallicity. Using these distances (their Table 4), the absolute magnitudes of the cluster variables in each cluster follow directly from their observed apparent magnitudes.

The result is a non-linear $M_V$, [Fe/H] relation (their Figs. 2, 3, and 4) with the opposite sense of the curvature than that of equation (12), as shown in Fig. 3. The non-linear equation put through the mean points in their Fig. 2a is,

$$M_V(RR) = 1.576 + 1.068 ([Fe/H]) + 0.242 ([Fe/H])^2. \qquad (15)$$

This is the CCMR line in Figure 3.

The difference in the sign of the curvature between equation (15) and our equation (12) is serious. By working the method of section 3 backwards through the pulsation equation (11) by again using the temperature from equation (6) and again equation (10) for the masses, both put into (11), and by using the same bolometric correction as in section 4.1, we obtain the period-metallicity relation that is required to produce the Caputo et al. eq. (15) to be,

$$\log P = -0.741 - 0.369 ([Fe/H]) - 0.081 ([Fe/H])^2, \qquad (16)$$

shown in Fig.4 as the dashed line.

The disagreement between the CCMR and the envelope log P-metallicity lines in Fig. 4 is not trivial because the CCMR line clearly violates the data in Figs. 1 and 2.

Three possible observational reasons, and at least one theoretical possibility exist to explain the reversal of the sign of the curvature between equations (12) and (15).

(1) Consider a possible systematic error in the CCMR calibration with [Fe/H] at the level of 0.2 mag due to (a) a population incompleteness bias, or (b) uncertainties in the calculated theoretical zero points in the crucial $M_V$, log P lines in the CCMR formulation.

(a) The CCMR method of fitting period-apparent magnitude data to derive cluster distances assumes that the shortest period c type variable in a given cluster defines the blue first harmonic edge. The vertical fitting of the data to the *theoretical* $M_V$, log P loci for the first harmonic blue edge is sensitive to a population incompleteness bias due to small-number statistics for the Bailey c-type variables in any given cluster. Obviously, a way to detect such a bias error, if it exists, is to compare the CCMR distances with independently determined distances by other methods, and to correlate the differences with the number of c-type variables used. The effect need only be at the 0.2 mag level to change the sense of the curvature in equation (15). CCMR have made part of this comparison in their Fig. 2, using the independent $M_V$([Fe/H]) calibrations of Fernley (1993), Fernley et al. (1998), and Carretta et al. (2000). None of these show a turn down in $M_V$ for [Fe/H] < -1.5.

(b). The possibility must be kept in mind that the CCMR theoretical calculation of the zero points of the period, absolute magnitude loci in the crucial $M_V$/log P diagram for various metallicites may be sensitive enough to the input physics that the CCMR fitting method may be in error at the 0.2 mag level in $M_V$. It relies on vertical displacements along $M_V$/log P lines of relatively steep slope, demanding that the calculated change of the zero point of these lines with metallicity (their equation 3) are correct at this level.

(2) However, in defense of the sense of the Caputo et al. curvature, we note again (emphasized by Caputo et al.) that most of the modern *theoretical* HB models (cf. Lee, DeMarque, & Zinn 1990; Castellani, Chieffi, & Pulone 1991; Bencivenni et al 1991; Dorman 1992, Caputo et al. 1993) predict a non-linear $M_V$/metallicity relation that has the *same* sign as the CCMR curvature. Why then is the disagreement of the CCMR line in log P-[Fe/H] correlation of Fig. 4 with the observational data in Figs. 1 and 2 so blatant?

The theoretical predictions of the HB models just cited are for the *age zero* horizontal branches, whereas the stars on the parabolic locus in Fig. 1 for [Fe/H] < -1.8 may be appreciably evolved on average, (they have longer observed periods than a linear locus would predict at these low metallicities), passing through the instability strip at brighter luminosities than the AZHB locus on their way to the AGB.

Clearly, the effect of evolution is a *brightening* of $M_V$ relative to the AZHB at fixed [Fe/H]. Hence, one could suppose that the curvature upward of line (1) in Fig. 3 (eq. 12) is not a property of the *age zero* HB, but may be due to luminosity evolution at the lowest metallicites due to the different morphologies of the HB as a function of [Fe/H]. This is the suggestion of Lee, Demarque, and Zinn (1990), but note the caveat by Rood (1990).

Nevertheless, this explanation does not account for the *opposite* sense of the non-linearity between equations (12) and (15) because the CCMR cluster data are also affected by luminosity evolution. Therefore, if this explanation of the sense of the curvature in equation (12) is correct, then the CCMR calibration itself would still have to be systematically incorrect with varying [Fe/H] at the 0.2 mag level, if our equation (12) is correct.

(3). But we can also criticize the sense of the upturn in equation (12). Our adopted colour at the blue edge, putatively giving the temperature of the stars at the envelope lines in Figs. 1 and 2 (eq. 6), may not be systematically correct at the 0.03 mag level in the colours for all [Fe/H]. If we were to vary the colour of stars along the envelope line in Fig. 1 by 0.01 mag at [Fe/H] = -2.0 relative to the colour for higher metallicities, and 0.02 mag at [Fe/H] = -2.5, the upturn of line 1 in Fig. 3 would disappear. Indeed, with these changes, there would be a *downturn* starting at [Fe/H] = -2.0. Such a small variation of the blue envelope colour with metallicity along the envelope line is too small to determine reliably with the available colour data for the Layden field star data. It is at the limit of the normal determinations of the reddening corrections and the methods of obtaining "static star" mean color over the cycle. Hence the determination of more precise temperatures along the lines in Figs. 1 and 2 at the level of $\Delta \log T < 0.004$ is the extant stringent requirement to advance this pulsation method.

(4). Alternatively, our turn up in equation (12) could be eliminated if the "constant" of 13.687 in equation (11) itself would be a function of [Fe/H].

## 6. CALIBRATION OF THE LUMINOSITY ZERO POINT AT BASED ON RR LYRAES IN THE LMC

Although, remarkably, the theoretical calibrations of $M_V$(RR) discussed here (Fig. 3) all agree in their theoretical *predicted* zero points to within 0.10 mag over the metallicity range from [Fe/H] of 0.0 to -2.0, despite the criticisms of the last section, we clearly need a purely observational determination of the luminosity zero point.

Baade's (1952, 1956) plan to ultimately determine <$M_V$(RR)> by comparing to the P-L relation of long period classical Cepheids in galaxies where both are seen, is now becoming a reality. This is the modern application of the method invented by Shapley (1918).

RR Lyraes have been discovered in many local galaxies whose distances are known either from classical Cepheids, or now, by other means; (Cepheid distances are presently in question at the 0.2 mag level because of the current demonstrable non-unique P-L relation from galaxy-to-galaxy; Tammann et al. 2003, Sandage et al. 2004).

Photometry for a large sample of RR Lyraes in LMC by Clementini et al. (2003a) gives $M_V$(RR) = + 0.52 at [Fe/H] = -1.5 when their data are combined with an LMC distance modulus of $(m - M)_o$ = 18.54, adopted by them, which also is the *non-Cepheid* value adopted by Tammann et al. (2003) from many independent methods. The Clementini et al. calibration is plotted as a cross in Fig. 3. Its agreement with equations (12), (13), and (15) is excellent, differing from them by only 0.04, 0.03, and 0.02 mag respectively at the stated metallicity. Note also that the sensitivity to metallicity in the Clementini et al. calibration (the slope) is 0.214 $\pm$ 0.047, which agrees with all the equations in sections 4 and 5 to within the errors. It is also to be emphasized that the Clementini et al. empirical zero point is the mean luminosity of the *evolved* horizontal branch because the LMC variables are at the mean evolution HB level, not the AZHB.

Improved zero point and slope calibrations, and a test for uniqueness from galaxy to galaxy, can be expected with this method by eventually combining the results from other galaxies in which RR Lyraes have been found (Dolphin et al. 2001 for IC 1613; Fusi Pecci et al. 1996, Clementini et al. 2001a, and Rich et al. 2001 for M31; Clementini et al. 2001b for the Leo I dwarf; Clementini et al. 2003b for NGC 6822 where a more complete literature listing is given).

## 7. SUMMARY OF VARIOUS VALUES OF THE $dM_V/d([Fe/H])$ SLOPE COEFFICIENT

### 7.1 Representative linear approximations to $M_V([Fe/H])$

There are at least a score of modern calibrations of $M_V(RR)$ as a function of [Fe/H]. It remains for a review article to summarize them all. Here we cite a representative sample that cover the range of the slope, $b$, in the linear approximation of $M_V = a + b$ ([Fe/H]).

Equation (1), from S93b, is among the calibrations with the largest value of $b = 0.30$. Its slope and zero point of $a = 0.94$ are closely supported by McNamera (1997a,b) who derives $M_V(RR) = 0.96 + 0.29$ ([Fe/H]) from his application of the Baade-Becker-Wesselink method. He also obtained $M_V(RR) = 1.00 + 0.31$([Fe/H]) using RR Lyraes with different metallicities in the Galactic bulge using data by Alcock et al. (1998), and $M_V(RR) = 1.06 + 0.32$ ([Fe/H]) (McNamera 1999) in his rediscussion of Fernley's BBW results. Feast (1997) derives $M_V = 1.13 + 0.37$ (Fe/H]) in discussing the bias problems in Fernley's application of the BBW method.

Smaller values of $b$ are more common. Fernley (1993) derived $b = 0.19$ using $(V-R)_o$ colors. In assessing the BBW radial velocity correction to the center of the stellar disk, Fernley (1994) later obtained $b = 0.21$. Fernley et al. (1998) again derived $b = 0.18$ from new BBW data. Carretta et al. (2000) obtained $b = 0.18$, principally from main sequence fittings.

From a study of eight clusters in M31 that have RR Lyrae photometric data, Fusi Pecci et al. (1996) obtained a shallow slope of $b = 0.13$. Later they revised the value to $b = 0.22$ from a larger sample of M31 clusters (Rich et al. 2001).

The most secure determination to date is the aforementioned work of Clementini et al. (2003a) from their large LMC sample where they derive $b = 0.214 \pm 0.047$ with no indication of an appreciable non-linearity with [Fe/H].

The smallest value of the slope was $b = -0.03$ by Jones, Carney, and Latham (1988) who used their early BBW results to state "that there is little if any dependence of $M_V$ upon metallicity" from their data that gave $<M_V(RR)> = 0.82$ at $<[Fe/H]> = -0.6$ and 0.87 mag at $<[Fe/H]> = -2.2$. They neglected the Oosterhoff effect which had earlier proved to be the clue leading to the initial model explaining the Oosterhoff period ratios that required $M_V(RR)$ to be a significant function of metallicity (S58, S81a,b). Jones et al. (1992) and Carney et al. (1992) did later revise their assessment using a larger BBW sample, obtaining $b = 0.16$, which however is smaller than that derived by McNamera (1999), also using BBW data which gave $b = 0.32$, consistent at the time with $b = 0.30$ derived from the model and the steep slope of the metallicity-period correlation itself (i.e Figs. 1 & 2 here, or Figs. 1-11 of S93a).

*7.2 Non-linear calibrations*

As mentioned above, early indications from theoretical AZHB models suggested that the variation of $M_V$ with [Fe/H] is non-linear (cf. Lee et al. 1990); Castellani et al. 1991; Bencivenni et al. 1991; Dorman 1992; Caputo et al. 1993).

In addition, papers by Caloi, D'Antona, & Mazzitelli (1997), Cassissi et al. (1998), and Demarque et al. (2000) again posited that the slope for age zero HB variables may not be the same for [Fe/H] above and below [Fe/H] = -1.5.

Caputo et al. (1993) had earlier suggested that $b$ depends on HB morphology in a way similar to the discussion in section 5. In addition, Caputo (1997) had shown from models that $b = 0.19$ for metal poor clusters with [Fe/H] < -1.5 and $b = 0.32$ for metal richer clusters with [Fe/H] > -1.5. This is consistent with the sense of the non-linearity of equation (15) which gives $b = 0.10$ at [Fe/H] = -2, 0.34 at -1.5, and 0.58 at -1, in the opposite sense from our calibration of equation (12) for reasons likely to reside in the demonstrable luminosity evolution away from the age zero HB (S90a; Sandage, Dietheim, & Tammann 1994, Figs. 1-4).

## 8. SUMMARY AND CONCLUSIONS

There are seven principal research results in this paper:

(1) A non-linear (parabolic) envelope at the shortest period for fundamental mode variables in the log P, [Fe/H] diagram of Fig. 1 (equation 3) replaces the linear approximation of Eq. (2) that was used in S93a,b.

(2) Assuming that stars on this envelope are at the fundamental blue edge of the instability strip, and that the colour at that edge is $(B-V)_o = 0.24$ for all metallicites, requires that the temperature of the envelope stars depends on [Fe/H] by equations (6) to (8) for various colour/temperature calibrations.

(3) Entering the pulsation equation (11), and using the temperature/colour scale in equation (6), the log P-[Fe/H] envelope from equation (3), and the mass-metallicity relation in equation (10) gives the non-linear $M_V$(RR),[Fe/H] relation of equation (12). The luminosity variation with [Fe/H] shows an upward turn relative to a linear increase at the lowest metallicities. The sense of the non-linearity can be reversed if we assume that the colour of the blue-edge variables for [Fe/H] more metal poor than -1.7 becomes redder by ~ 0.03 mag from that of more metal rich variables. New precision colour and reddening data are required to test the possibility.

(4) Alternatively, if equation (3) is replaced by the linear log P-[Fe/H] envelope of equation (4), the predicted $M_V$(RR) calibration is the linear equation (13) whose slope is $dM_V/d([Fe/H]) = 0.18$ and whose zero point is $M_V = 0.82$.

(5) The non-linear $M_V$(RR) calibration of Caputo et al. (2000) in equation (15) has the opposite sign of the curvature with [Fe/H] than equation (12), although it agrees with model calculations by a number of groups for the age zero horizontal branches. However, Fig. 4 shows that the sense of the CCMR non-linearity in their $M_V$ calibration is in blatant disagreement with the *observed* log P, [Fe/H] data in Figs. 1 and 2 that are central to the problem. Suggestions in section 5 for a solution include (a) the effect of evolution from the age zero horizontal branch on the theoretical loci in a $M_V$, log P diagram used by CCMR to derive their semi-empirical $M_V$(RR) values, and (b) to question our assumption concerning the colour to be used to obtain the needed temperatures for the envelope stars, given the high sensitivity to temperature in the pulsation equation, which will cause our sense to be incorrect.

(6) However, the consequences of the disagreement between equations (12) and (15) are only slight over the metallicity range of -1.0 to -2.5, as displayed in Fig. 3. The values of the slope, $b = dM_V/d([Fe/H])$ embrace the entire range of the values in the recent literature, albeit with the opposite sign for the [Fe/H] variation between equations (12) and (15). Both show the slope to range between $b = 0$ to 0.6 in the following way. From equation (12) from [Fe/H] = 0.0 to -2.5, the $dM_V/d([Fe/H])$ slopes are (-0.04, 0.06, 0.16, 0.26, 0.35, and 0.45) for [Fe/H] values of (0.0, -0.5, -1.0, -1.5, -2.0, and -2.5). For the CCMR equation (15) between [Fe/H] of -1.0 and -2.5, the slopes are (0.58, 0.34, 0.10, and -0.14) for metallicites of (-1.0, -1.5, -2.0, and -2.5).

(7) The zero points of $M_V$, based solely on the adopted (semi-theoretical) zero points in equations ((3), (6), (10), and (11) agree to within 0.1 mag between equations (12), (15) and the new linear envelope line in equation (4) over the metallicity range from -1.0 to -2.5, as shown in Figure 3.

However, our calibration here using the above equations may or may not refer to the age zero horizontal branch, depending on whether the envelope lines in the log P-metallicity relations in Fig. 1 and 2 and the adopted temperature-metallicity relation of equation (6) do or do not refer to the age zero blue fundamental edges, but rather to an average RR Lyrae ensemble that has evolved from the age zero configuration.

To circumvent the problem, we automatically include the effect of evolution by normalizing the theoretical $M_V$ lines in Fig. 3 by using the *observed* zero point of $M_V = +0.52$ at [Fe/H] = -1.5 determined by Clementini et al. (2003a) from their LMC data that clearly represent the average evolved HB state. With this empirical normalization, we adopt as a compromise calibration,

$$M_V = + 0.82 + 0.20 \,([Fe/H]), \qquad (17)$$

which is almost identical to the calibration by Fernley (1993) based on $(V-R)_o$ colours using an independent method, and is within 0.08 mag of equation (1) over the metallicity range from -0.5 to -2.0, as derived in S93a. These clearly support the "long" distance scale.

## ACKNOWLEDGMENTS

It is a pleasure to thank Leona Kershaw for preparing the text from many intermediate drafts, written in a powerful but now obscure word editor program, and David Sandage for the preparation of the diagrams in digital form for publication.

FIGURE CAPTIONS

**Figure 1.** The log period-metallicity relation for 301 Bailey type ab RR Lyraes in the general field from the data by Layden (1994). The adopted envelope that represents the shortest period variables at a given period is the parabolic equation (3) of the text. Note the four outriders at shorter periods than the envelope, identified in the text.

**Figure 2.** Same data as in Fig. 1 but with the most extreme linear envelope at the shortest periods, as a solid line, that can accommodate the two stars AA Aql and SS Gru. It is equation (4) of the text. The dashed line is the ridge line through the mean of the data, binned in intervals of [Fe/H] with normal points from Table 1. It is equation (5) of the text.

**Figure 3**. Summary of the three calibrations using a pulsation equation as it is combined with the adopted correlations of period, temperature, and mass variations with [Fe/H] from equations (3), (4), (6), (10), and (11). Line 1 is equation (12), line 2 is equation (13), and the CCMR line is from equation (15). The observed empirical point, shown as a cross at $M_V = + 0.52$, [Fe/H] = - 1.5, is from Clementini et al.(2003a) from LMC RR Lyrae data.

**Figure 4**. The observed period-metallicity blue envelope lines from Figs. 1 and 2 whose equations are (3) and (4), plus the inferred relation, marked CCMR, that is required if the Caputo et al. (2000) calibration of equation (15) is used in the pulsation equation (11), solved backwards using equations (15), (10), and (6) as input.

TABLE 1

THE MID-POINT RIDGE LINE OF THE PERIOD-METALLICITY CORRELATION IN FIGURE 1

| Item | <[Fe/H]> | Range of [Fe/H] | | n | <log P> | <B-V>$_o$ |
|---|---|---|---|---|---|---|
| (1) | (2) | (3) | | (4) | (5) | (6) |
| mean | -0.359 | +0.07 | -0.70 | 21 | -0.378 | 0.340 |
| rms | 0.049 | | | | 0.015 | 0.011 |
| mean | -0.983 | -0.71 | -1.19 | 16 | -0.329 | 0.335 |
| rms | 0.039 | | | | 0.011 | 0.013 |
| mean | -1.468 | -1.20 | -1.79 | 71 | -0.268 | 0.327 |
| rms | 0.019 | | | | 0.007 | 0.005 |
| mean | -1.991 | -1.80 | -2.49 | 34 | -0.217 | 0.329 |
| rms | 0.031 | | | | 0.012 | 0.007 |
| mean | -2.278 | -2.06 | -2.49 | 7 | -.200 | 0.331 |
| rms | 0.059 | | | | 0.018 | 0.025 |

# TABLE 2

## RESIDUALS IN LOG $T_e$ BETWEEN THE SBT TEMPERATURE SCALE AND THREE OTHERS*

| [Fe/H] | 0.00 | -0.50 | -1.00 | -1.31 | -1.66 |
|---|---|---|---|---|---|
| (1) | (2) | (3) | (4) | (5) | (6) |
| B-V | | | | | |
| (LOG $T_e$; SBT MINUS SEKIGUCHI & FUKUGITA | | | | | |
| 0.2 | -.012 | .000 | +.002 | +.006 | +.010 |
| 0.3 | -.007 | +.002 | .000 | +.004 | +.005 |
| 0.4 | -.006 | -.001 | -.007 | -.003 | -.003 |
| 0.5 | -.009 | -.006 | -.011 | -.008 | -.006 |
| 0.6 | ---- | -.011 | -.013 | -.010 | -.014 |
| 0.7 | ---- | -.005 | -.012 | -.008 | ---- |
| (Log $T_e$; SBT MINUS KOVACS & WALKER | | | | | |
| 0.2 | +.013 | +.016 | +.011 | +.011 | +.014 |
| 0.3 | +.016 | +.016 | +.007 | +.007 | +.008 |
| 0.4 | +.017 | +.013 | -.002 | .000 | +.001 |
| 0.5 | +.018 | +.011 | -.011 | -.002 | -.001 |
| 0.6 | ---- | +.011 | +.002 | +.001 | -.004 |
| 0.7 | ---- | +.024 | +.011 | +.010 | ---- |
| (Log $T_e$; SBT MINUS DORMAN | | | | | |
| 0.2 | ---- | +.017 | +.005 | +.005 | +.003 |
| 0.3 | ---- | +.015 | +.007 | +.006 | +.010 |
| 0.4 | ---- | +.013 | +.008 | +.004 | +.006 |
| 0.5 | ---- | +.008 | +.003 | +.002 | +.006 |
| 0.6 | ---- | +.003 | -.002 | -.001 | -.003 |
| 0.7 | ---- | +.006 | -.004 | -.001 | ---- |

* Plus sign means that SBT is hotter than others.

TABLE 3

THE SLOPE OF THE TEMPERATURE- METALLICITY CORRELATION
AT FIXED (B-V)o FOR FOUR INDEPENDENT CALIBRATIONS*

| (B-V)o | SBT | SF | KW | Dormaan |
|---|---|---|---|---|
| (1) | (2) | (3) | (4) | (5) |
| 0.20 | (.007) | (.018) | .007 | .001 |
| 0.24 | .010 | .018 | .007 | .007 |
| 0.30 | .015 | .020 | .007 | .010 |
| 0.33 | .016 | .021 | .007 | .011 |
| 0.40 | .018 | .021 | .007 | .013 |
| 0.50 | .020 | .019 | .007 | .016 |
| 0.60 | .021 | .021 | .007 | .011 |
| 0.70 | (.027) | .022 | .007 | .017 |

*Listed is the slope, $d\log T_e/d[Fe/H]$, at fixed colours

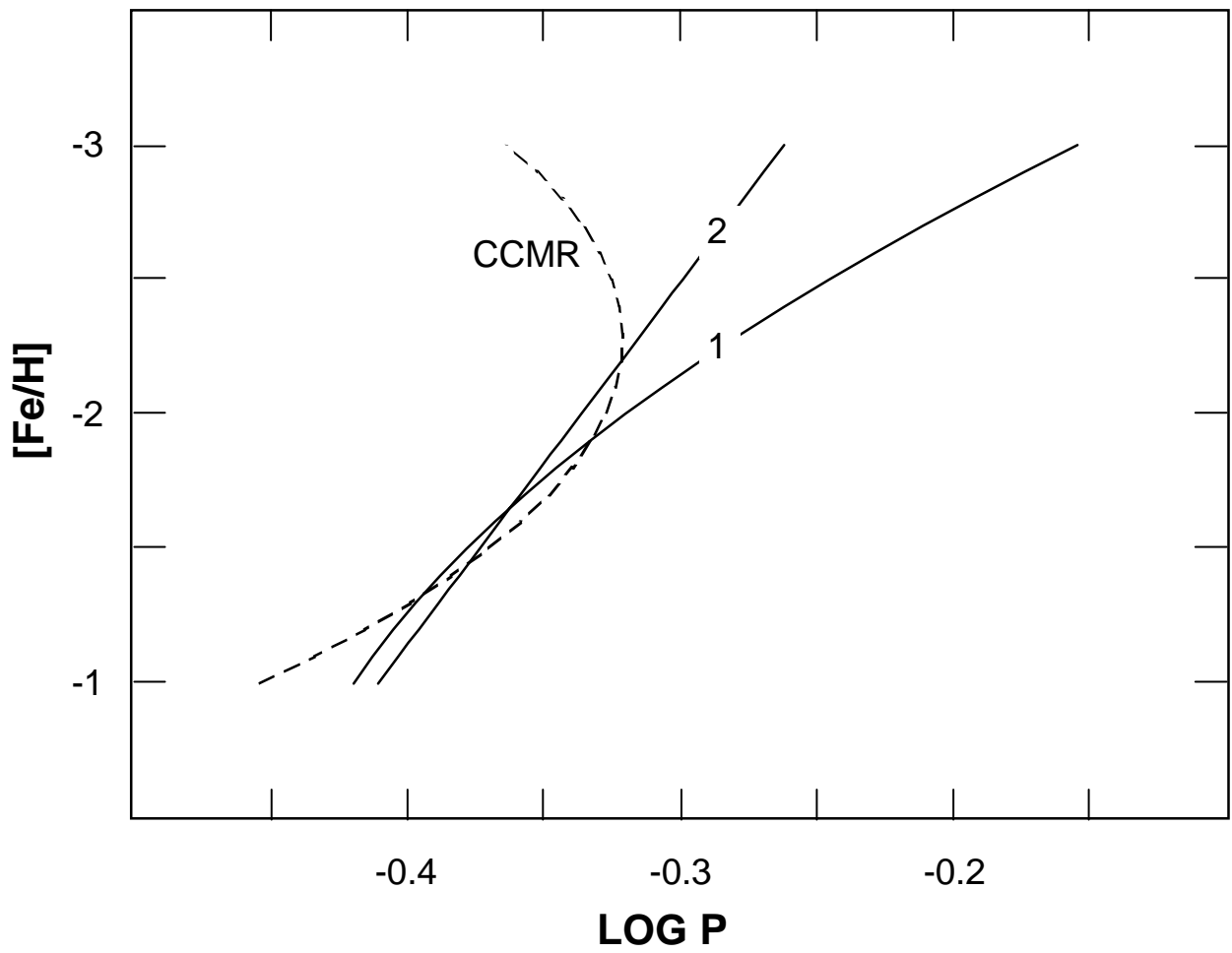

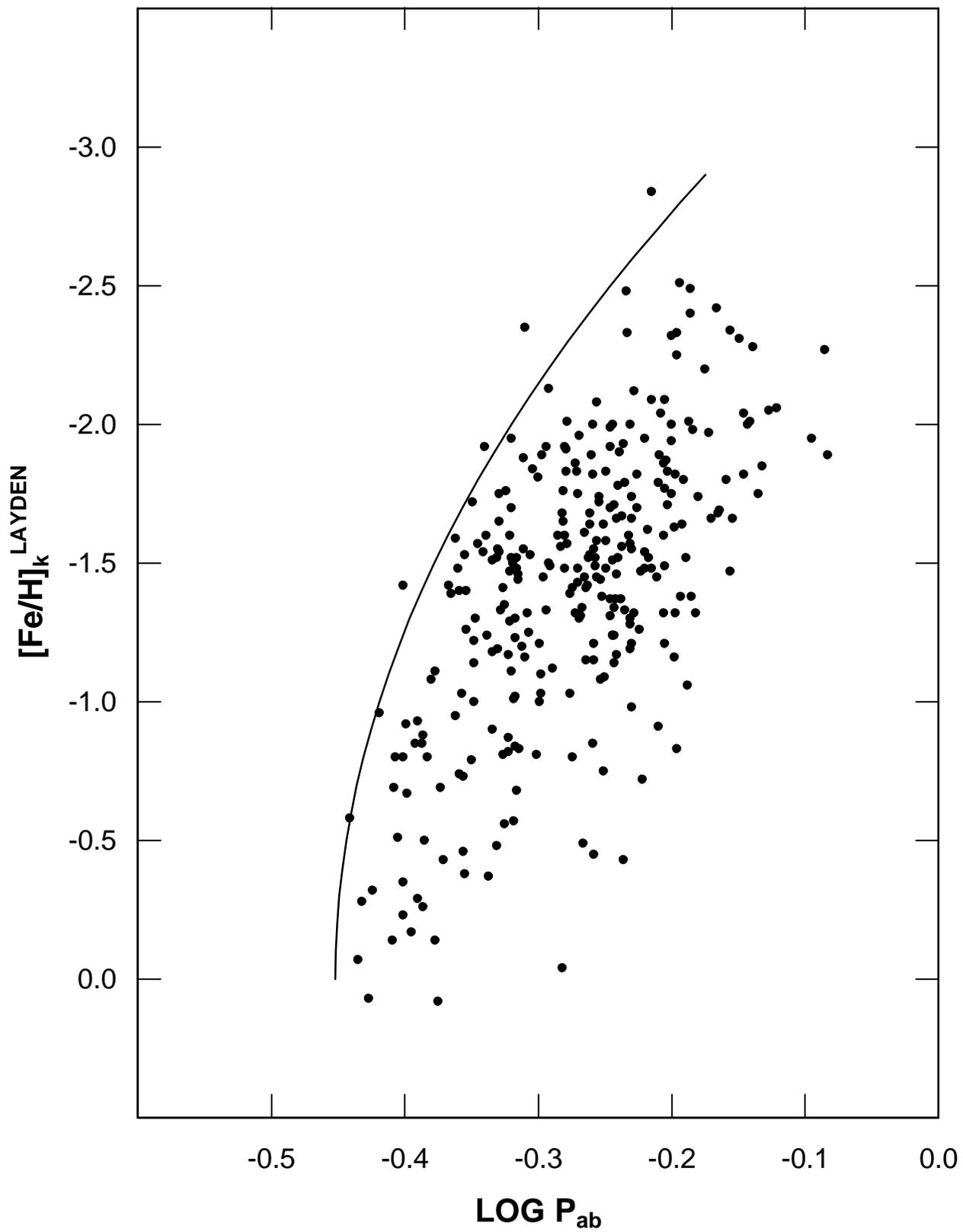

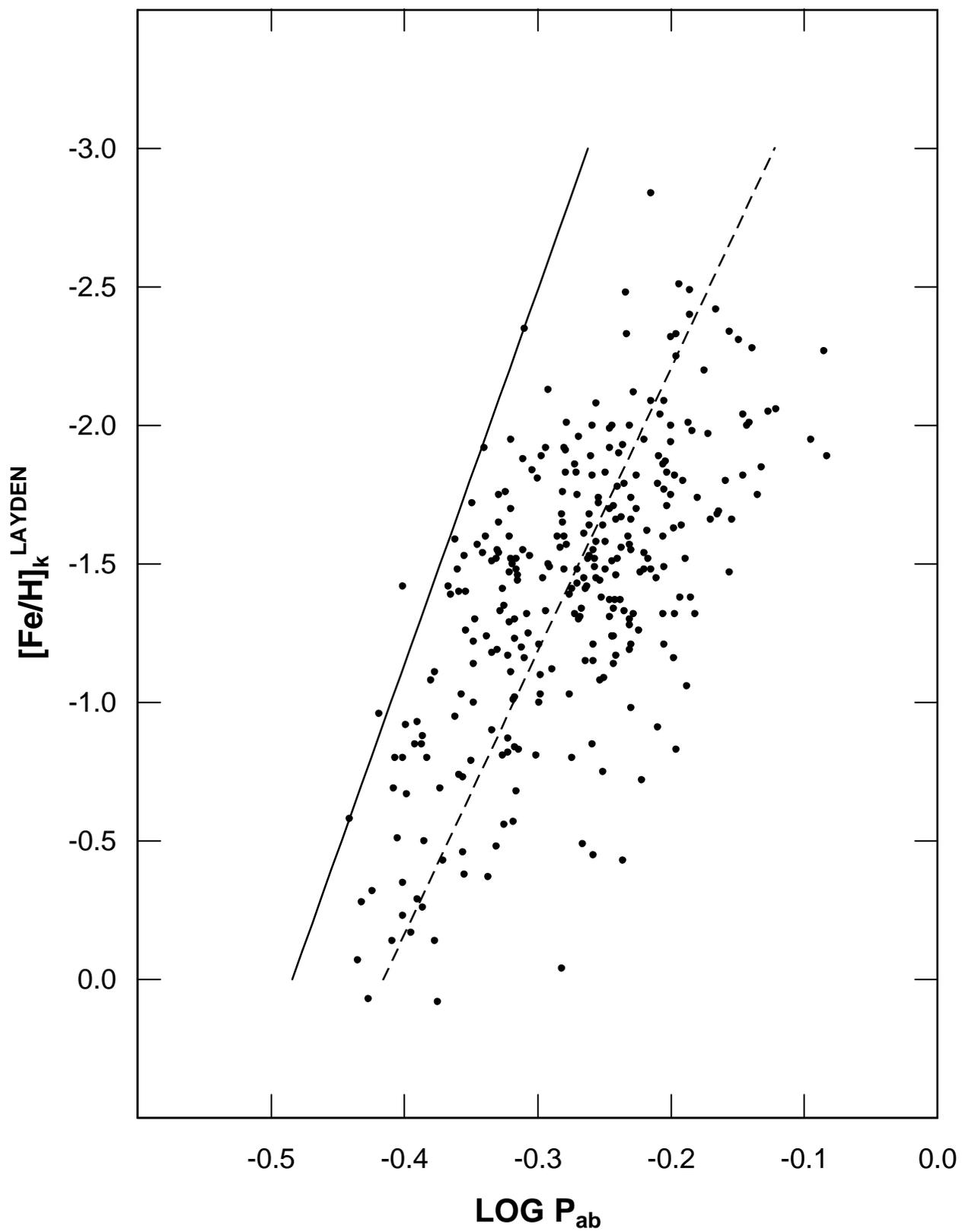

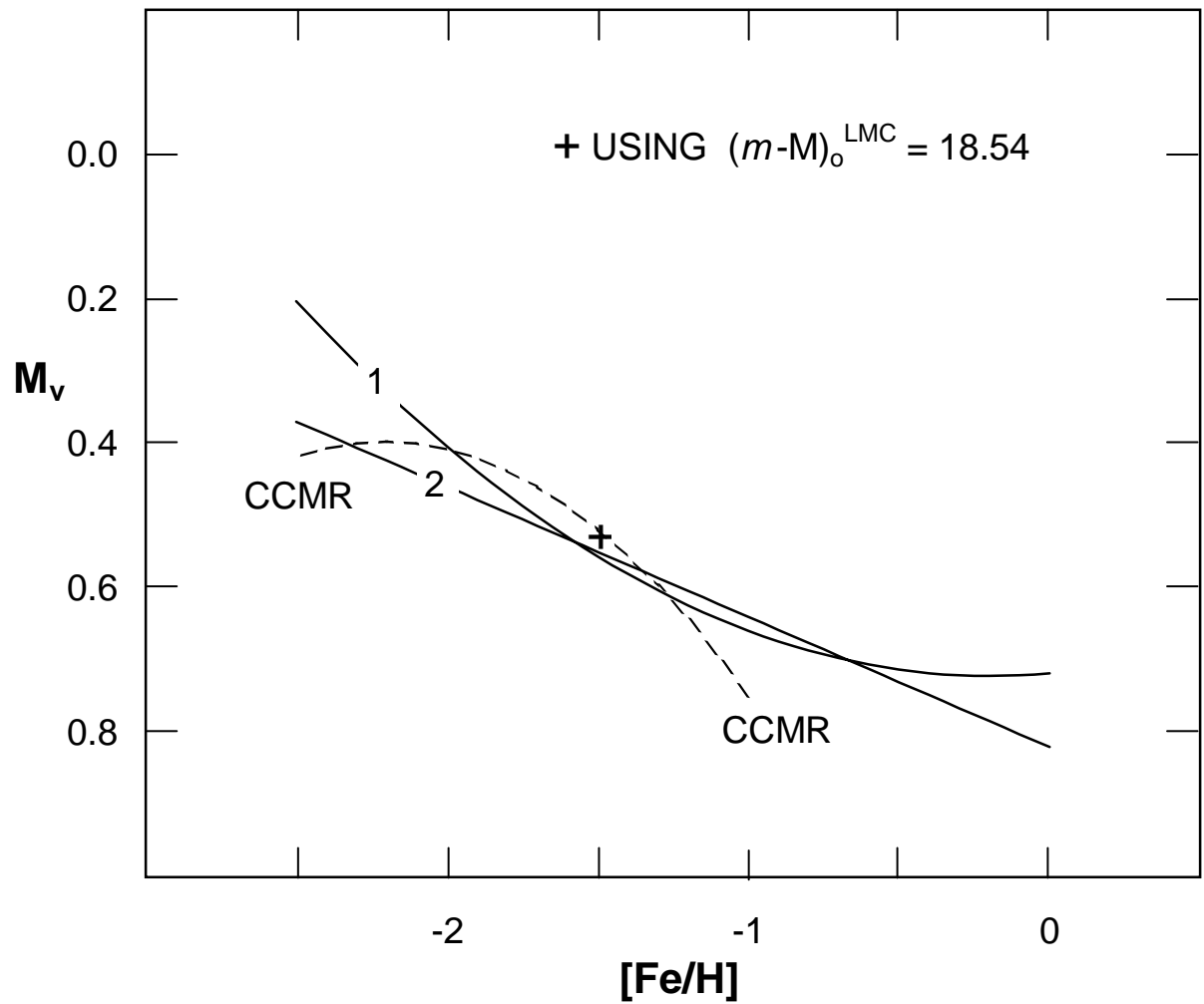